\documentclass[journal]{IEEEtran}

\usepackage{amsmath,amssymb,amsfonts}
\usepackage{newtxmath}
\usepackage{booktabs}
\usepackage{cite}
\usepackage{graphicx}
\usepackage{multirow}
\usepackage{float}
\usepackage{algorithm}
\usepackage{algorithmic}
\usepackage{textcomp}
\usepackage{xcolor}
\usepackage{url}
\usepackage[hidelinks]{hyperref}

\newcommand{\prevrev}[1]{#1}
\newcommand{\rev}[1]{#1}

\graphicspath{{figures/}}

\title{Uncertainty-Guided Counterfactual Sensor Selection and Safety-Gated Recognition for Autonomous Sensing Agents in IoT Systems}
\author{Shirong~Guo, Hongming~Zhou, Muhammad Attique Khan, Jinfeng Xu, Yonis Gulzar, Manar Mousa Altamimi, and Kunfu~Wang%
\thanks{Shirong Guo is with the Faculty of Engineering, Monash University, Clayton, Victoria 3800, Australia (e-mail: shirong.guo@monash.edu).}%

\thanks{Muhammad Attique Khan is with the Center of AI, Prince Mohammad bin Fahd University, Alkhobar, Saudi Arabia (e-mail: attique.khan@ieee.org)}

\thanks{Jinfeng Xu is
with the Department of Electrical and Computer Engineering, The University
of Hong Kong, Hong Kong, China (e-mail: jinfeng@connect.hku.hk).}

\thanks{Yonis Gulzar is with the Department of Management Information Systems, College of Business Administration, King Faisal University, Al-Ahsa 31982, Saudi Arabia (e-mail: Ygulzar@kfu.edu.sa)}

\thanks{Manar Mousa Altamimi is with the Department of Information Systems, College of Computer and Information Sciences (CCIS), Princess Nourah bint Abdulrahman University, P.O. Box 84428, Riyadh 11671, Saudi Arabia (e-mail: mmaltamimi@pnu.edu.sa)}

\thanks{Hongming Zhou and Kunfu Wang are with the School of Advanced Manufacturing, Fuzhou University, Quanzhou 362251, China (e-mail: zhouhongming2025@126.com; okonfu23@gmail.com).}%
\thanks{Hongming Zhou and Kunfu Wang are the corresponding authors.}}

\begin{document}
\maketitle

\begin{abstract}
\prevrev{\rev{Autonomous IoT systems rely on distributed sensing, but sensor faults and communication loss can leave multichannel evidence incomplete or unreliable. Their agents must therefore classify measurements, assess evidence sufficiency, select a sensor to reacquire, and hand off unsafe decisions.} We formulate counterfactual sensing: a degraded window defines the factual state, each candidate restoration defines a controlled alternative, and a pre-request utility model estimates loss reduction. Physics-Guided Invariant-Kinematic Fusion (PGIK-Fusion) constructs class evidence, while calibrated risk and cost-sensitive utility select acceptance, sensor request, or handoff from factual information.} On the Logistic Activity Recognition Challenge dataset, PGIK-Fusion achieves 0.768 accuracy and a macro-averaged F1 score (Macro-F1) of 0.696, exceeding extreme gradient boosting (XGBoost)--one-dimensional convolutional neural network (1D-CNN) fusion by 2.49 and 2.00 percentage points, respectively. \prevrev{Across seven predefined degradation conditions on held-out subjects,} the policy reduces unsafe acceptance from 33.0\% to 13.8\%; relative to an action-rate-matched random policy, it raises automation accuracy by 18.5 percentage points and lowers expected cost by 24.7\%. \prevrev{In the \rev{auxiliary within-event wind-turbine study}, PGIK-Fusion reaches 0.975 Macro-F1, improves on XGBoost--1D-CNN by 1.17 percentage points, and reduces unsafe acceptance from 20.1\% to 8.0\%. \rev{The evaluation characterizes the controlled responses of a frozen recognition model to specified sensor restorations under predefined fault families and severities.} Within this scope, the results demonstrate that calibrated recognition, targeted evidence acquisition, and conservative deferral can form a unified perception--action loop that reduces unsafe automation under sensor degradation.}
\end{abstract}


\section{Introduction}

\IEEEPARstart{A}{n} intelligent agent is a computational entity situated in an environment, capable of perceiving that environment and selecting actions with a degree of autonomy to pursue its objectives \cite{wooldridge_1995_intelligent_agents,duan2025llm,huang2025nexus,xu2026agents,zhang2026hera,zhang2026memmark,duan2025leed}. Sensor systems instantiate this perception--action abstraction when measurements influence what, where, or when to sense next. Recent uncertainty-aware autonomous sensing in the Internet of Things (IoT), for example, constructs a sensing agent whose policy adapts measurement placement to an application objective \cite{murad_2024_autonomous_sensing}. This progression moves sensing beyond passive telemetry, but leaves open how an agent should manage degraded observations in multichannel recognition.

\begin{figure}[!t]
\centering
\includegraphics[width=\columnwidth]{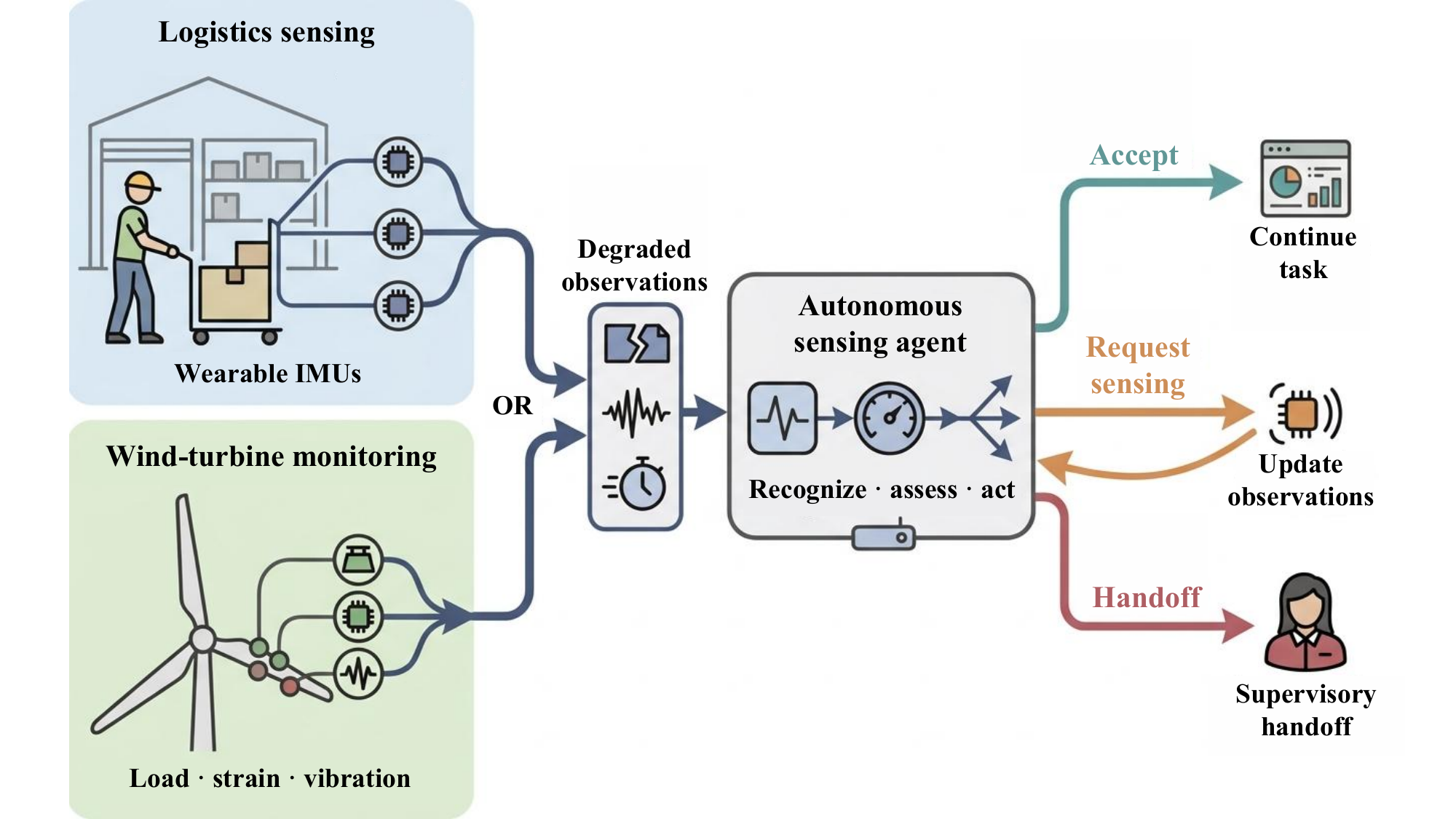}
\caption{Representative operating scenarios for evaluating the autonomous sensing agent in logistics activity recognition and wind-turbine monitoring.\label{fig:scenario-overview}}
\end{figure}

Here, an \emph{autonomous sensing agent in an IoT system} maps networked observations to three predefined actions: accept the recognition result, request one physical sensor, or hand the decision to an upper-level controller. Its \emph{task-bounded} autonomy is constrained by a recognition objective, the current information state, and this finite action set, rather than by an application domain. \emph{Safety-gated recognition} releases a factual label only after acceptance; a request delays release until \rev{one reacquisition and one model recomputation, after which the updated label is released directly}, while handoff withholds the automated label. The state combines class evidence, sensor quality, predictive uncertainty, and action costs. This abstraction applies to logistics, industrial monitoring, connected transportation, and autonomous vehicles \cite{bianchi_2019_personalized_har,yang_2022_secure_sensor_fusion}. Fig.~\ref{fig:scenario-overview} depicts the two representative applications evaluated here.

Missing channels, noise, packet loss, and delay can alter both the predicted class and the reliability of its evidence; predictive uncertainty may also deteriorate under dataset shift \cite{ovadia_2019_uncertainty_shift}. Clean-input accuracy therefore neither establishes confidence after degradation nor identifies which sensor could correct an error. Although uncertainty-aware IoT sensing can price additional measurements \cite{lee_2026_cost_efficient_sensing}, recognition still requires candidate-specific rules for reacquisition, continued automation, and handoff.

We address this gap by separating factual perception from counterfactual evidence control. Physics-Guided Invariant-Kinematic Fusion (PGIK-Fusion) combines raw temporal encoding with physical descriptors adapted to each sensing system. Controlled missingness, noise, delay, and packet loss produce paired model responses: the degraded input defines the factual state, and restoration of one requestable sensor defines a candidate state. A utility model fitted on validation data predicts each candidate's reduction in classification loss from factual evidence alone. Calibrated error risk, predicted benefit, and acquisition cost then determine acceptance, one sensor request, or handoff. The estimand is the frozen recognition model's response to a specified restoration.

Evaluation uses the Logistic Activity Recognition Challenge (LARa) version 03 inertial measurement unit (IMU) benchmark and \rev{an auxiliary within-event study using} the Multimodal Dataset for Wind Turbine Blade Monitoring during Lightning Strikes (MDWTBM-LS) \cite{niemmann_2020_lara,niemmann_2023_lara_v03,li_2025_mdwtbm}. Models are retrained with domain-specific descriptors and requestable modalities in each regime.

The main contributions are:
\begin{enumerate}
\item A model-level counterfactual formulation that represents factual degradation and candidate sensor restorations as paired model responses with an explicit boundary to natural causal identification.
\item A physics-guided perception backbone that combines coordinate-robust and kinematic descriptors with raw temporal convolution.
\item A cost-sensitive policy that estimates candidate-specific recovery from pre-request evidence and selects acceptance, one sensor request, or handoff.
\item A task-bounded agent evaluation across human- and equipment-centered sensing instances, connecting recognition, uncertainty, controlled intervention, action allocation, unsafe acceptance, and expected cost.
\end{enumerate}

\section{Related Work}

\subsection{Autonomous Sensing Agents and Multichannel Perception}

Agent-based sensing connects observations and internal state to objectives and actions \cite{wooldridge_1995_intelligent_agents,murad_2024_autonomous_sensing}. In air-quality and indoor-noise monitoring, objective-driven sensing increased critical-period detection by 78.5\% and 357.3\% while reducing energy use by 14.3\% and 7.6\%, respectively \cite{murad_2024_autonomous_sensing}; sensor-specific recovery, recognition risk, and handoff remain unresolved.

\prevrev{Deploying such agents near data sources introduces latency, bandwidth, privacy, and device-heterogeneity constraints \cite{lin2024split,fang2026aggregation,lin2025leo,qu2025mobile}. Adaptive and hierarchical split federated learning address resource allocation during distributed training through model partitioning and device-specific adaptation \cite{lin2024adaptsfl,tan2026exploiting,wei2025optimizing,lin2025hierarchical,fang2026nsc}. Inference-time sensor recovery instead requires an explicit decision over physical observations.}

\prevrev{Wearable activity recognition research has examined sensor placement, feature construction, subject variation, personalization, single-sensor efficiency, and modern sequence modeling \cite{hammerla_2016_deep_wearables,bianchi_2019_personalized_har,lu_2020_efficient_har,li_2025_p2lhap,li_2025_harmamba,yuan2024satsense,fang2025dynamic,yuan2026neural}. These methods improve recognition or efficiency but do not prescribe actions for unreliable sensors.}

\prevrev{LARa provides synchronized inertial and semantic data, whereas MDWTBM-LS provides blade load, strain, and vibration \cite{niemmann_2020_lara,niemmann_2023_lara_v03,li_2025_mdwtbm}. Published LARa results vary across modalities, partitions, and metrics \cite{niemmann_2020_lara,mekruksavanich_2023_resnext_lara}; they do not couple recognition to reacquisition and deferral.}

\subsection{Temporal and Physically Structured Representations}

Multichannel time-series benchmarks span tree ensembles such as Random Forest, extreme gradient boosting (XGBoost), and the Light Gradient Boosting Machine (LightGBM) \cite{breiman_2001_random_forests,chen_2016_xgboost,ke_2017_lightgbm}, together with raw-sequence encoders based on long short-term memory (LSTM), the gated recurrent unit (GRU), self-attention, and causal dilated convolution \cite{hochreiter_1997_lstm,cho_2014_gru,vaswani_2017_attention,bai_2018_tcn}. These families provide broad task coverage but do not specify which physical sensor should be restored after degradation.

Complementary representations encode temporal geometry through Gramian angular fields \cite{wang_2015_imaging_timeseries}, while TimesNet, ModernTCN, and TimeMIL model periodicity, cross-variable interaction, and time-aware instance evidence \cite{wu_2023_timesnet,luo_2024_moderntcn,chen_2024_timemil}. In other recognition modalities, Interpretable Fixed-Length Representation for Fingerprint Matching via Vision Transformer (IFViT) combines global and local fixed-length representations, whereas Vision-Context Hierarchical Attention Autoregressive With Adaptive Permutation (HAAP) couples visual and contextual evidence. Graph and physics-guided methods additionally introduce sensor relations or domain structure \cite{kipf_2017_gcn,willard_2022_scientific_knowledge}. These approaches enrich factual representations but do not estimate the prospective value of restoring a degraded physical sensor.

\subsection{Uncertainty and Intervention-Aware Sensing}

Risk-aware automation requires confidence that tracks empirical correctness. Temperature scaling and shift-aware evaluation provide complementary evidence about calibration quality \cite{guo_2017_calibration,ovadia_2019_uncertainty_shift}. Temperature scaling reduced expected calibration error from 12.67\% to 0.96\% for a 110-layer residual network on the Canadian Institute for Advanced Research 100 dataset \cite{guo_2017_calibration}, while distribution-shift experiments showed that clean-data calibration need not persist after input change \cite{ovadia_2019_uncertainty_shift}. Calibration does not identify which sensor would improve a decision.

Selective prediction controls risk and coverage through rejection or confidence ranking \cite{elyaniv_2010_selective,geifman_2019_selectivenet,cattelan_2024_confidence}; SelectiveNet reduced mean coverage violation from 11.98 to 3.63 percentage points \cite{geifman_2019_selectivenet}. Conformal methods provide finite-sample set coverage under exchangeability \cite{shafer_2008_conformal_tutorial,angelopoulos_2023_conformal}. Both assess current evidence, whereas sensor recovery requires prospective utility.

\subsection{Counterfactual Reasoning and Cost-Aware Acquisition}

Counterfactual foundations distinguish observed evidence from responses under specified alternatives through potential outcomes and identifiability conditions \cite{rubin_1974_causal_effects,pearl_2019_seven_tools,scholkopf_2021_causal_representation}. Algorithmic recourse further requires feasible interventions \cite{karimi_2021_algorithmic_recourse}. The present estimand is the fixed recognition model's controlled response to a specified restoration; identifying natural failure effects would require a separate causal design.

Active feature acquisition converts information value into cost-sensitive actions through reinforcement learning, variational representations, or generative surrogates \cite{janisch_2019_costly_features,ma_2019_eddi,li_2021_active_feature}; IoT soft sensing also links uncertainty to operating cost \cite{lee_2026_cost_efficient_sensing}. The present task restores one degraded physical sensor, supervises candidate-specific loss reduction, and retains handoff when neither current nor restored evidence supports automation.

\section{Methodology}

The framework formulates task-bounded autonomous sensing as a finite evidence-management loop: PGIK-Fusion constructs class evidence, calibration characterizes reliability, controlled restorations supply candidate-utility targets, and a cost-sensitive gate selects acceptance, one request, or handoff.

\begin{figure*}[!t]
\centering
\includegraphics[width=\textwidth]{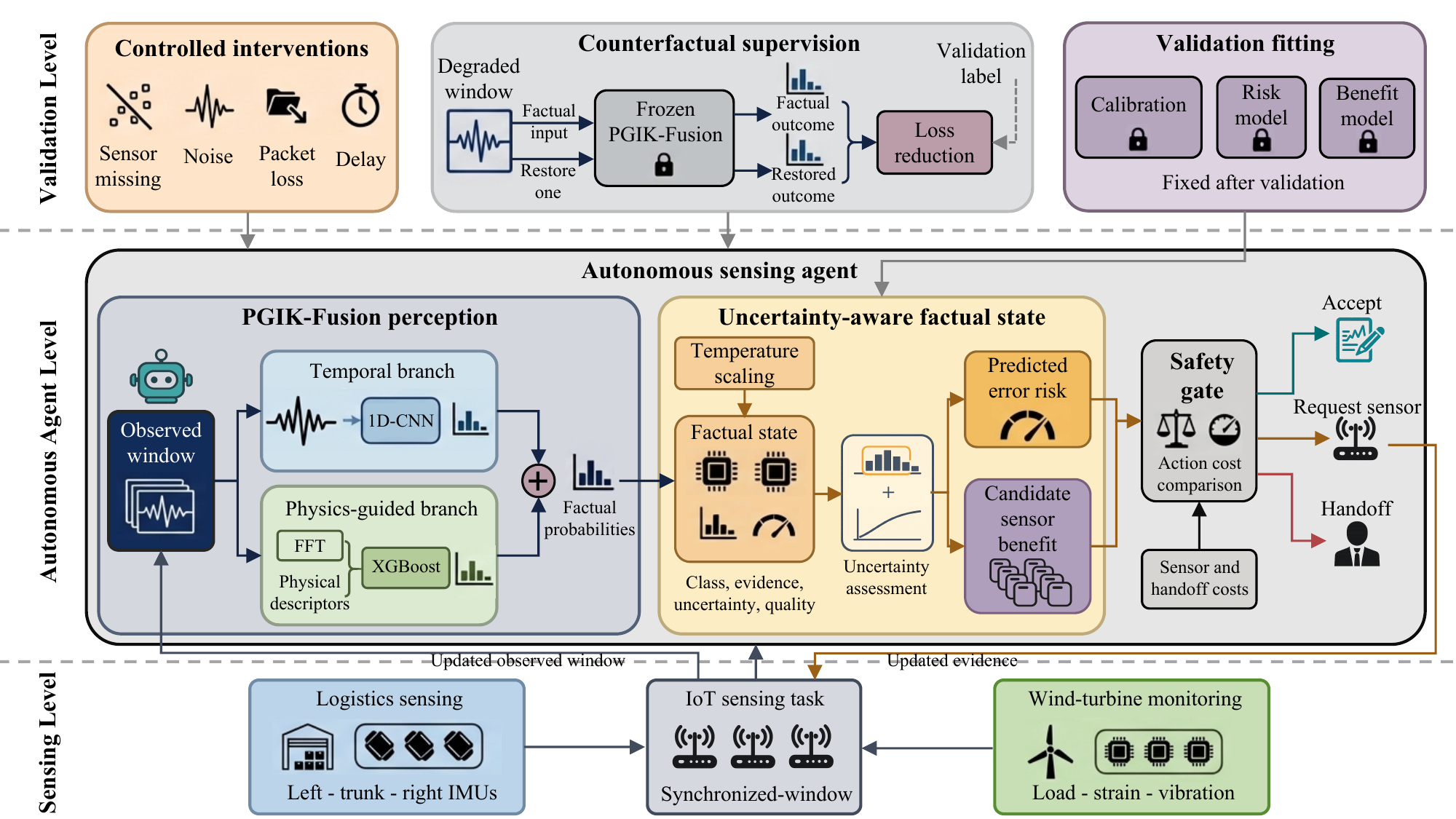}
\caption{End-to-end framework for uncertainty-guided counterfactual sensor selection and safety-gated recognition. The upper, middle, and lower layers denote validation, autonomous sensing agent operation, and representative evaluation instantiations, respectively.\label{fig:framework-overview}}
\end{figure*}

Fig.~\ref{fig:framework-overview} separates validation, autonomous agent operation, and domain-specific sensing while identifying the components fixed before deployment.

\prevrev{At validation, frozen responses to factual and single-sensor-restored windows define targets for calibration, risk, and benefit models. At deployment, pre-request evidence determines factual risk, candidate benefit, and action cost. Accordingly, the deployed selector uses neither restored observations nor labels, and all fitted components remain fixed during test evaluation. A request reacquires or retransmits the selected source and recomputes affected features; acceptance releases the prediction, whereas handoff invokes the fallback. The fast Fourier transform (FFT) icon denotes the spectral component in Section~III-B.}

\subsection{Task Formulation and Requestable Sensors}

The decision problem is defined at the level of one synchronized window so that every action is tied to an explicit information state. Let $x_t^{(0)}\in\mathbb{R}^{L\times D}$ contain $L$ samples from $D$ channels in the observed, possibly degraded state, and let $y_t\in\{1,\ldots,K\}$ be its label. Superscript $(0)$ denotes this factual model input. The frozen perception model produces the uncalibrated class vector $\boldsymbol{\pi}_t^{(0)}$ and its associated prediction as
\begin{equation}
\left\{\begin{aligned}
\boldsymbol{\pi}_t^{(0)}&=p_f(\cdot\mid x_t^{(0)})\in[0,1]^K,\\
\hat y_t^{(0)}&=\operatorname*{arg\,max}_{1\leq k\leq K}\pi_{tk}^{(0)}.
\end{aligned}\right.
\label{eq:factual-prediction}
\end{equation}
Here $p_f$ is the frozen fusion classifier and $\pi_{tk}^{(0)}$ is the uncalibrated probability of class $k$. Let $\mathcal V=\{v_j\}_{j=1}^{J}$ contain the $J$ requestable sensors. The feasible action set is $\mathcal A=\{\mathrm{accept},\mathrm{handoff}\}\cup\{\mathrm{request}(v_j)\mid v_j\in\mathcal V\}$. Acceptance immediately releases $\hat y_t^{(0)}$; \rev{a request postpones release until candidate $v_j$ is reacquired and the updated window is evaluated once, after which the resulting label is released directly without a second gate;} handoff suppresses automated release and invokes a conservative fallback. This release rule defines safety-gated recognition independently of the perception architecture.

The pre-request state must distinguish predictive ambiguity from deficient sensor availability. It therefore concatenates the calibrated class vector $\mathbf p_t^{(0)}$ defined in Section~III-D with three complementary uncertainty summaries and the sensor-quality vector:
\begin{equation}
\mathbf z_t=\left[(\mathbf p_t^{(0)})^\top,c_t,H_t,m_t,\mathbf q_t^\top\right]^\top
\in\mathbb R^{K+3+J},
\label{eq:agent-state}
\end{equation}
where the scalar summaries are defined separately as
\begin{equation}
c_t=\max_k p_{tk}^{(0)},
\label{eq:confidence}
\end{equation}
\begin{equation}
H_t=-\sum_{k=1}^{K}p_{tk}^{(0)}\log p_{tk}^{(0)},
\label{eq:entropy}
\end{equation}
\begin{equation}
m_t=p_{t(1)}^{(0)}-p_{t(2)}^{(0)}.
\label{eq:margin}
\end{equation}
These summaries encode distinct aspects of decision reliability. Equations (\ref{eq:confidence})--(\ref{eq:margin}) quantify peak confidence, distributional ambiguity, and separation between the two leading classes, respectively, with the convention $0\log 0=0$. Here $p_{t(1)}^{(0)}$ and $p_{t(2)}^{(0)}$ are the largest and second-largest calibrated probabilities, while $\mathbf q_t\in[0,1]^J$ records the quality of the $J$ requestable sensors. Retaining the full probability vector distinguishes windows with equal confidence but different class evidence.

The sensing configuration defines the request set. For LARa, $\mathcal V=\{v_l,v_t,v_r\}$ denotes the left, trunk, and right IMUs; for MDWTBM-LS, the candidates are load, strain, and vibration. Each candidate is restored independently, and the decision layer contains no sensor graph.

\subsection{Physics-Guided Invariant-Kinematic Fusion}

PGIK-Fusion combines temporal representation learning with explicitly structured evidence through two independently trained probability estimators. The temporal branch maps the standardized multichannel window to class probabilities through a one-dimensional convolutional neural network (1D-CNN). The physics-guided branch concatenates statistical--spectral descriptors with sensing-system-specific physical descriptors before XGBoost classification. The branches do not exchange latent features; their probability vectors are combined only by the convex fusion rule defined below. This separation preserves the interpretation of the structured descriptors while allowing the temporal branch to learn local waveform patterns.

The structured branch first establishes a domain-neutral statistical--spectral basis so that raw channels and physically derived sequences can be summarized on the same scale. For a scalar sequence $s=(s_1,\ldots,s_L)$, define
\begin{equation}
\left\{
\begin{aligned}
\boldsymbol{\psi}(s)
&=\left[(\boldsymbol{\psi}^{\mathrm{stat}}(s))^{\top},
(\boldsymbol{\psi}^{\mathrm{spec}}(s))^{\top}\right]^{\top},\\
\boldsymbol{\psi}^{\mathrm{stat}}(s)
&=\left[\mu_s,\sigma_s,\min_{1\leq\tau\leq L}s_\tau,
\max_{1\leq\tau\leq L}s_\tau,\right.\\
&\qquad\left.Q_{0.1}(s),Q_{0.5}(s),Q_{0.9}(s),
\operatorname{RMS}(s)\right]^{\top},\\
\boldsymbol{\psi}^{\mathrm{spec}}(s)
&=\left[\bar A_1(s),\bar A_2(s),\bar A_3(s),\bar A_4(s)\right]^{\top}.
\end{aligned}
\right.
\label{eq:summary-operator}
\end{equation}
Here $\mu_s$, $\sigma_s$, $Q_q(s)$, and $\operatorname{RMS}(s)$ denote the sample mean, standard deviation, empirical $q$-quantile, and root-mean-square value. The term $\bar A_b(s)$ is the mean discrete-Fourier-transform magnitude in the $b$th of four equal nonnegative-frequency index bands, where $b\in\{1,2,3,4\}$. The operator therefore retains location, dispersion, range, distribution shape, signal energy, and coarse frequency allocation without requiring a domain-specific frequency boundary. Applied to the 27 LARa channels, (\ref{eq:summary-operator}) produces 324 base descriptors.

The physical augmentation then introduces invariances that are not guaranteed by channel-wise summaries. This construction follows theory-guided learning by exposing domain-structured evidence without imposing a complete governing equation \cite{karpatne_2017_theory_guided}. Let $\mathbf u_\tau^{(s)}\in\mathbb R^3$ denote the triaxial signal at sample $\tau$ for modality $s\in\{a,\omega,m\}$, corresponding to acceleration, angular velocity, and magnetic field. Orientation-insensitive magnitude and first-order kinematic change are defined as
\begin{equation}
\rho_{\tau}^{(s)}=\left\lVert\mathbf u_{\tau}^{(s)}\right\rVert_2,
\qquad s\in\{a,\omega,m\},
\label{eq:magnitude}
\end{equation}
\begin{equation}
\delta_{\tau}^{(s)}=
\left\lVert\mathbf u_{\tau}^{(s)}-\mathbf u_{\tau-1}^{(s)}\right\rVert_2,
\qquad s\in\{a,\omega\}.
\label{eq:kinematic}
\end{equation}
Directional consistency complements the two norm-based operators by retaining relative orientation:
\begin{equation}
\kappa_{\tau}=\frac{\mathbf a_{\tau}^{\top}\mathbf m_{\tau}}
{\lVert\mathbf a_{\tau}\rVert_2\lVert\mathbf m_{\tau}\rVert_2+\epsilon}.
\label{eq:direction}
\end{equation}
\prevrev{Equation~(\ref{eq:magnitude}) applies for $\tau=1,\ldots,L$, and (\ref{eq:kinematic}) for $\tau=2,\ldots,L$. Here $\mathbf a_\tau=\mathbf u_\tau^{(a)}$, $\mathbf m_\tau=\mathbf u_\tau^{(m)}$, and $\epsilon=10^{-6}$ prevents instability when either vector norm is near zero. The operators $\rho_\tau^{(s)}$, $\delta_\tau^{(s)}$, and $\kappa_\tau$ represent magnitude, first-order change, and directional consistency, respectively. At each IMU, the six derived sequences are summarized by (\ref{eq:summary-operator}), adding 216 descriptors that encode axis-robust magnitude, transition strength, and relative direction.}

The invariant terminology follows from the behavior of these operators under a shared change of orthonormal sensor coordinates. For any $\mathbf R\in\mathbb R^{3\times3}$ satisfying $\mathbf R^\top\mathbf R=\mathbf I$, where $\mathbf I$ is the $3\times3$ identity matrix,
\begin{equation}
\left\{
\begin{aligned}
\left\lVert\mathbf R\mathbf u_\tau\right\rVert_2
&=\left\lVert\mathbf u_\tau\right\rVert_2,\\
\left\lVert\mathbf R(\mathbf u_\tau-\mathbf u_{\tau-1})\right\rVert_2
&=\left\lVert\mathbf u_\tau-\mathbf u_{\tau-1}\right\rVert_2,\\
\kappa(\mathbf R\mathbf a_\tau,\mathbf R\mathbf m_\tau)
&=\kappa(\mathbf a_\tau,\mathbf m_\tau).
\end{aligned}
\right.
\label{eq:rotation-invariance}
\end{equation}
Here $\kappa(\cdot,\cdot)$ denotes the normalized inner-product operator in (\ref{eq:direction}). Equation~(\ref{eq:rotation-invariance}) establishes invariance to a common orthogonal rotation or reflection of the triaxial coordinate frame. It does not assume invariance to independent sensor displacement or time-varying misalignment, which remain represented through the observed sequences.

The complete LARa feature map makes the relation between these operators and the reported dimensionality explicit. Let $\mathcal L=\{l,t,r\}$ index the left, trunk, and right IMUs, and let $\mathcal S_{\mathrm{phys}}(x)$ collect the 18 derived sequences: $\rho_{\ell}^{(s)}$ for $s\in\{a,\omega,m\}$, $\delta_{\ell}^{(s)}$ for $s\in\{a,\omega\}$, and $\kappa_\ell$, at every $\ell\in\mathcal L$. The physics-guided representation is
\begin{equation}
\boldsymbol{\phi}_{\mathrm{PGIK}}(x)=\operatorname{concat}\!\left(
\{\boldsymbol{\psi}(x_d)\}_{d=1}^{27},
\{\boldsymbol{\psi}(s)\}_{s\in\mathcal S_{\mathrm{phys}}(x)}
\right)\in\mathbb R^{540}.
\label{eq:pgik-feature-map}
\end{equation}
In (\ref{eq:pgik-feature-map}), $x_d$ is the $d$th channel and $\operatorname{concat}$ denotes concatenation; the raw and derived blocks contain $27\times12=324$ and $18\times12=216$ descriptors. Their shared summary operator isolates the effect of physical structure without changing the classifier family.

\prevrev{The equipment-sensing instantiation preserves the representation principle while changing its physical primitives. For MDWTBM-LS, load, strain, and vibration at two positions form six modality--position vectors. Their magnitudes and first differences yield 12 sequences, while three cosine-similarity sequences compare positions within each modality. Summarization expands the 144-dimensional statistical--spectral representation to 324 dimensions. The shared construction encodes within-vector dynamics and cross-position agreement, while the wearable and structural sensor semantics remain domain specific.}

Independent branch training preserves structured and temporal evidence without forcing a shared latent space. XGBoost \cite{chen_2016_xgboost} uses 250 trees, depth 6, learning rate 0.08, and row and feature subsampling of 0.85. Class $k$ receives weight $\sqrt{N/n_k}$, where $N$ and $n_k$ are the total and class-$k$ training counts, moderating rare-class weights. Following direct convolutional wearable encoding \cite{ordonez_2016_deepconvlstm}, the temporal branch uses 64, 128, and 192 channels with kernel widths 7, 5, and 5. The first two layers apply batch normalization, a Gaussian error linear unit, and max pooling; the third precedes global average pooling and a linear classifier. Inputs are standardized using training-subject statistics only.

Probability fusion provides an explicit balance between structured and temporal evidence. For input window $x$ and class $y$, let $\pi_p(y\mid x)$ and $\pi_c(y\mid x)$ denote the uncalibrated physics-guided XGBoost and raw 1D-CNN probabilities, respectively. Their fused output is
\begin{equation}
\pi_f(y\mid x)=\lambda \pi_p(y\mid x)+(1-\lambda)\pi_c(y\mid x).
\label{eq:fusion}
\end{equation}
The weight $\lambda\in[0,1]$ controls structured evidence, and convexity preserves a valid probability vector. Validation Macro-F1 selects $\lambda$ from $\{0,0.05,\ldots,1\}$, yielding 0.90 on LARa. Features, fitted branches, and $\lambda$ are then fixed before calibration, interventions, and testing; $p_f$ in (\ref{eq:factual-prediction}) denotes this predictor.

\subsection{Controlled Interventions and Sensor Utility}

Controlled interventions pair a degraded factual input with one restored candidate. Complete IMU loss zeros one nine-channel group; Gaussian noise uses 0.10 or 0.25 times the training-channel standard deviation; packet loss zeros trunk samples 80--119 (20\% of a 200-sample window); and delay shifts the right IMU by 10 samples with zero padding (0.10~s at 100~Hz). These conditions represent source removal, amplitude corruption, contiguous outage, and misalignment. Both branches and all derived features are recomputed after each intervention. \prevrev{The same predefined intervention families and severities are used for policy development and held-out-subject evaluation. The resulting evidence measures controlled robustness and action allocation within these conditions. Generalization to previously unseen fault mechanisms or severity levels requires separate evaluation.}

\prevrev{Sensor-quality metadata describe degradation without revealing the unavailable clean signal. Clean and missing sensors receive 1 and 0; packet loss and delay use the retained-sample fractions 0.80 and 0.95; moderate and strong noise use one minus the normalized noise amplitude, giving 0.90 and 0.75. These indicators populate $\mathbf q_t$ as condition-supplied inputs rather than field fault-detector outputs.}

\begin{figure}[!t]
\centering
\includegraphics[width=\columnwidth]{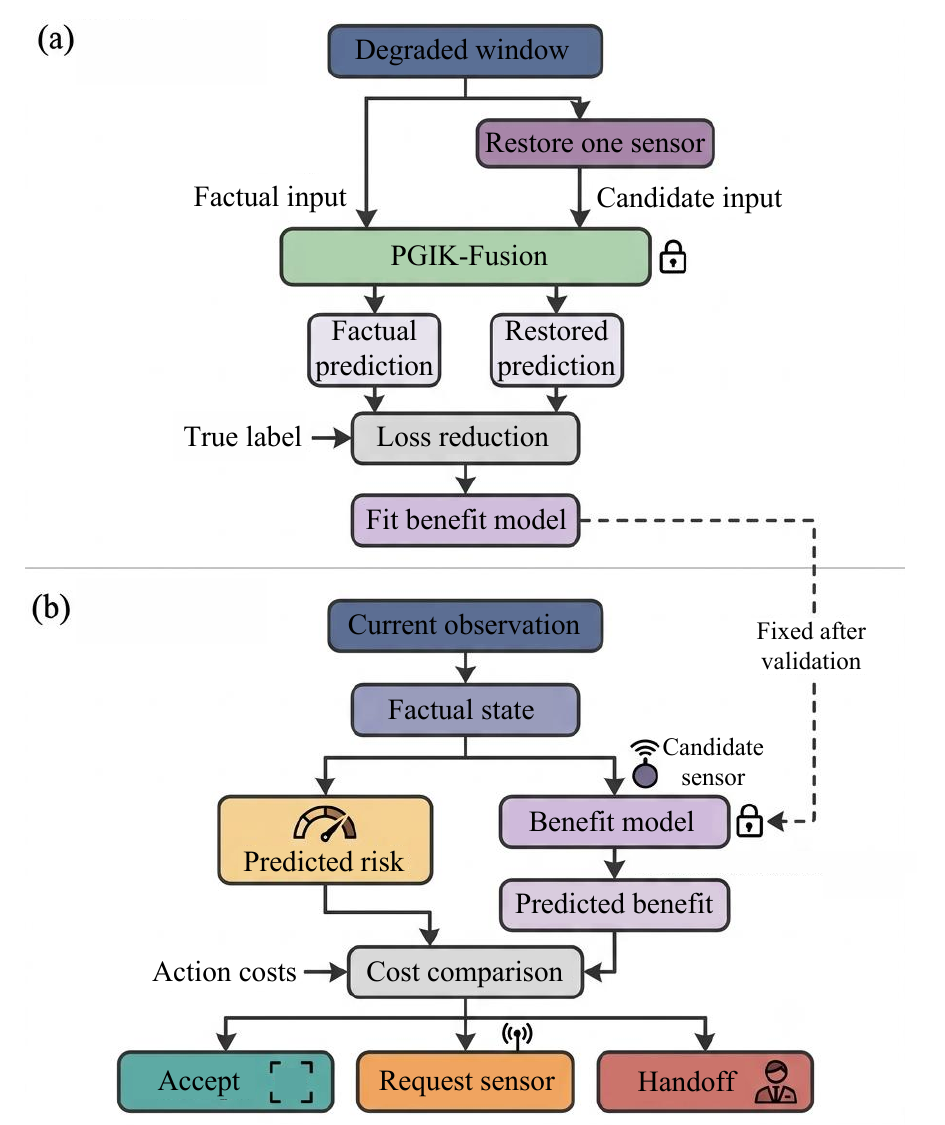}
\caption{Counterfactual sensor benefit and finite action selection: (a) validation-stage supervision from controlled restoration; (b) deployment using pre-request factual information.\label{fig:counterfactual-mechanism}}
\end{figure}

Candidate restoration defines a supervised counterfactual target while preserving the deployment information boundary. Let $\mathcal I_j$ restore candidate $v_j$ while leaving all other observed channels unchanged. The corresponding candidate state is
\begin{equation}
x_t^{(j)}=\mathcal I_j\!\left(x_t^{(0)}\right),\qquad j\in\{1,\ldots,J\}.
\label{eq:candidate-world}
\end{equation}
\prevrev{During target construction, the frozen classifier gives $\boldsymbol{\pi}_t^{(j)}=p_f(\cdot\mid x_t^{(j)})$ and $\hat y_t^{(j)}=\operatorname*{arg\,max}_{k}\pi_{tk}^{(j)}$. Equation~(\ref{eq:candidate-world}) changes exactly one requestable source, thereby isolating its realized contribution while holding all nonrequested channels fixed. Fig.~\ref{fig:counterfactual-mechanism} separates validation supervision, where restorations and labels are available, from deployment selection, where the policy receives only $\mathbf z_t$ and candidate identity $v_j$.}

The realized candidate benefit is defined on decision correctness rather than confidence change. For 0--1 classification loss $\ell(y,\hat y)=\mathbf 1[y\neq\hat y]$, where $\mathbf 1[\cdot]$ is the indicator function, the loss reduction under candidate $j$ is
\begin{equation}
\Delta\ell_{tj}=\ell\!\left(y_t,\hat y_t^{(0)}\right)-\ell\!\left(y_t,\hat y_t^{(j)}\right).
\label{eq:loss-reduction}
\end{equation}
\prevrev{The signed target equals 1 for error correction, 0 for unchanged correctness, and $-1$ for a newly introduced error. It is used only for target construction and retrospective analysis; neither $y_t$ nor $x_t^{(j)}$ is available before a request.}

The policy decomposes action value into factual error risk and candidate-specific recovery benefit because high risk alone does not identify which sensor should be requested. A standardized logistic model fitted on the S11--S12 validation subjects estimates
\begin{equation}
\hat r_t=\sigma\!\left(\beta_0+\boldsymbol{\beta}^{\top}\widetilde{\mathbf z}_t\right)
\approx \Pr\!\left(\hat y_t^{(0)}\neq y_t\mid\mathbf z_t\right),
\label{eq:risk-model}
\end{equation}
where $\sigma(u)=(1+e^{-u})^{-1}$ is the logistic link, $\widetilde{\mathbf z}_t$ is standardized using S11--S12 statistics, $\beta_0$ is the intercept, and $\boldsymbol{\beta}$ contains the fitted coefficients. The logistic model uses an $\ell_2$ penalty with inverse regularization strength 1 and a maximum of 1000 solver iterations; regularization stabilizes the estimate under correlated probability and uncertainty inputs. A standardized ridge model separately estimates the expected candidate benefit as
\begin{equation}
\hat b_{tj}=h_{\boldsymbol\theta}(\widetilde{\mathbf z}_t,v_j)
\approx\mathbb E\!\left[\Delta\ell_{tj}\mid\mathbf z_t,v_j\right].
\label{eq:expected-benefit}
\end{equation}
Here $h_{\boldsymbol\theta}$ is a ridge predictor using standardized state $\widetilde{\mathbf z}_t$, one-hot candidate identity, and their interactions, allowing one factual window to assign different sensor utilities. \prevrev{Its penalty is fixed at 10 before test evaluation to shrink the interaction-expanded coefficient vector under the sparse nonzero target. Predictions are then clipped to the target support $[-1,1]$ so that extrapolation cannot create an invalid loss reduction before cost evaluation.}

Resource cost converts a predicted recovery into a deployable acquisition decision. For candidate $v_j$, the normalized cost is
\begin{equation}
C_j=w_E\widetilde E_j+w_D\widetilde D_j+w_B\widetilde B_j.
\label{eq:resource-cost}
\end{equation}
Here $\widetilde E_j$, $\widetilde D_j$, and $\widetilde B_j$ denote normalized energy, latency, and bandwidth burden; the nonnegative weights satisfy $w_E+w_D+w_B=1$. Because the three LARa IMUs have equal channel widths and the dataset provides no platform telemetry, the primary experiment assigns $C_j=0.10$ to every request and evaluates sensitivity over 0.05--0.25. \prevrev{For a deployed platform, $\widetilde E_j$ can be estimated from sensor wake-up, sampling, and transmission telemetry; $\widetilde D_j$ from reacquisition, queueing, and transmission delay relative to the decision deadline; and $\widetilde B_j$ from payload and retransmission volume relative to the available link budget. Normalizing these measurements by device-specific budgets and selecting $w_E$, $w_D$, and $w_B$ from application priorities yields sensor- and context-dependent request costs without changing the decision rule.} The MDWTBM-LS experiment retains the same normalized request and handoff costs for load, strain, and vibration; its validation-based modality selection and recovery gate are specified in Supplementary Section S1.3.

\subsection{Uncertainty and Finite Safety Actions}

Probability calibration is required because the risk estimator and action loss use class probabilities as decision evidence rather than only as a class ranking. Temperature scaling preserves the predicted class while adjusting probability sharpness \cite{guo_2017_calibration}. For fused output $\boldsymbol{\pi}_t^{(0)}$, the calibrated probability of class $k$ is
\begin{equation}
p_{tk}^{(0)}(T)=
\frac{\left(\pi_{tk}^{(0)}\right)^{1/T}}
{\sum_{h=1}^{K}\left(\pi_{th}^{(0)}\right)^{1/T}},
\qquad T>0.
\label{eq:temperature-scaling}
\end{equation}
\prevrev{Here $h$ indexes the classes. The transform is equivalent to applying softmax to $\log\boldsymbol{\pi}_t^{(0)}/T$ and covers sharpening ($T<1$) and softening ($T>1$). The calibration temperature is selected on the validation partition by}
\begin{equation}
T^{\star}=\operatorname*{arg\,min}_{T\in\mathcal T}
-\frac{1}{|\mathcal D_{\mathrm{cal}}|}
\sum_{i\in\mathcal D_{\mathrm{cal}}}
\log p_{i y_i}^{(0)}(T),
\label{eq:temperature-objective}
\end{equation}
\prevrev{where $\mathcal D_{\mathrm{cal}}$ denotes S11--S12, $i$ indexes its windows, $y_i$ is the associated label, and $\mathcal T$ contains 101 equally spaced values from 0.5 to 3.0. LARa selects $T^{\star}=1.45$, and $\mathbf p_t^{(0)}$ henceforth denotes the calibrated vector used in (\ref{eq:agent-state})--(\ref{eq:margin}). Calibration is assessed by 15-bin expected calibration error (ECE), negative log-likelihood (NLL), multiclass Brier score, and ten confidence-ranked coverage levels from 0.10 to 1.00. Fifteen bins balance probability resolution against validation occupancy, whereas the coverage grid provides a compact trajectory from selective to complete automation.}

Set-valued uncertainty complements scalar calibration by retaining all labels compatible with a finite-sample threshold. For a scoring rule fixed independently of the calibration observations, the split-conformal rank construction targets marginal coverage under exchangeability without requiring a parametric error model \cite{shafer_2008_conformal_tutorial,angelopoulos_2023_conformal}. For $n_{\mathrm{cal}}$ calibration windows, define nonconformity scores $a_i=1-p_{i y_i}^{(0)}(T^{\star})$ and their order statistics $a_{(1)}\leq\cdots\leq a_{(n_{\mathrm{cal}})}$. The finite-sample threshold is
\begin{equation}
\hat q=a_{(r)},\qquad
r=\min\!\left\{n_{\mathrm{cal}},
\left\lceil(n_{\mathrm{cal}}+1)(1-\alpha)\right\rceil\right\},
\label{eq:conformal-quantile}
\end{equation}
and the prediction set for factual window $t$ is
\begin{equation}
\Gamma_t=\left\{k\in\{1,\ldots,K\}:
1-p_{tk}^{(0)}(T^{\star})\leq\hat q\right\}.
\label{eq:conformal-set}
\end{equation}
Here $\alpha$ is target miscoverage, $r$ the corrected order-statistic index, and $\hat q$ the fixed threshold. \prevrev{We set $\alpha=0.10$ to target 90\% marginal coverage, yielding $\hat q=0.933$ on LARa.} The conformal set complements scalar calibration but does not enter the action rule. Since S11--S12 support both temperature selection and quantile estimation, the observed test coverage is reported empirically; a disjoint conformal-calibration partition would be required for the standard finite-sample guarantee.

The finite safety gate completes the evidence-management loop by placing acceptance, acquisition, and deferral on one cost scale. Safety gating here denotes cost-sensitive suppression of high-risk automated outputs through sensor request or handoff:
\begin{equation}
\mathcal L_t(a)=
\begin{cases}
\hat r_t, & a=\mathrm{accept},\\
\operatorname{clip}(\hat r_t-\hat b_{tj},0,1)+C_j, & a=\mathrm{request}(v_j),\\
C_h, & a=\mathrm{handoff}.
\end{cases}
\label{eq:action-cost}
\end{equation}
The branches in (\ref{eq:action-cost}) share a unit because an emitted error has unit cost. Here $\operatorname{clip}(u,0,1)=\min\{1,\max\{0,u\}\}$ bounds residual risk; acceptance incurs factual risk, a request adds $C_j$ after predicted recovery, and handoff pays $C_h$. An optimal action satisfies
\begin{equation}
a_t^{\star}\in\operatorname*{arg\,min}_{a\in\mathcal A}\mathcal L_t(a).
\label{eq:optimal-action}
\end{equation}
The action boundary becomes explicit by separating recoverable risk from acquisition burden. Define
\begin{equation}
\left\{
\begin{aligned}
d_{tj}&=\hat r_t-
\operatorname{clip}(\hat r_t-\hat b_{tj},0,1),\\
U_{tj}&=d_{tj}-C_j,\\
\mathcal L_t(\mathrm{request}(v_j))&=\hat r_t-U_{tj}.
\end{aligned}
\right.
\label{eq:net-recovery-utility}
\end{equation}
Here $d_{tj}$ is the signed predicted change in factual error risk after restoring $v_j$; it is negative when restoration is predicted to increase risk. The quantity $U_{tj}$ is the corresponding net utility after acquisition cost. Let $j_t^\dagger\in\operatorname*{arg\,max}_{1\leq j\leq J}U_{tj}$ denote any maximizing candidate. A strict preference for requesting this candidate over both acceptance and handoff is equivalent to
\begin{equation}
U_{t j_t^\dagger}>
\max\left\{0,\hat r_t-C_h\right\}.
\label{eq:request-region}
\end{equation}
The first term in (\ref{eq:request-region}) requires predicted recovery to offset acquisition cost; the second requires the post-request loss to fall below the deferral cost. Thus, sensor identity is selected by candidate-specific net utility, while the release decision remains jointly controlled by factual risk and $C_h$. When action losses are equal, (\ref{eq:optimal-action}) represents the corresponding set of minimizing actions.

The cost ordering determines the remaining policy behavior without a hard confidence threshold. The handoff cost is $C_h=0.30$, placing conservative deferral above a single request cost but below the unit cost assigned to an emitted error. With $C_j=0.10<C_h<1$, low-risk observations can be accepted, recoverable high-risk observations can trigger acquisition, and high residual risk can trigger handoff. Let $g$ and $h$ denote the fitted risk and benefit models, respectively. Algorithm~\ref{alg:counterfactual-gate} summarizes the resulting deployment procedure.

\begin{algorithm}[H]
\caption{Counterfactual Sensor Selection and Safety Gating}
\label{alg:counterfactual-gate}
\begin{algorithmic}[1]
\REQUIRE Factual window $x_t^{(0)}$, quality $\mathbf q_t$, frozen PGIK-Fusion model $p_f$, risk model $g$, benefit model $h$, costs $\{C_j\}_{j=1}^{J}$ and $C_h$
\STATE Compute $\mathbf p_t^{(0)}$, $\hat y_t^{(0)}$, and pre-request state $\mathbf z_t$
\STATE Estimate $\hat r_t\leftarrow g(\mathbf z_t)$
\FOR{$j=1$ to $J$}
\STATE Estimate $\hat b_{tj}\leftarrow h(\mathbf z_t,v_j)$ without observing $x_t^{(j)}$
\STATE Evaluate $\mathcal L_t(\mathrm{request}(v_j))$ using (\ref{eq:action-cost})
\ENDFOR
\STATE Evaluate acceptance and handoff losses; select a minimizer $a_t^\star$ from (\ref{eq:optimal-action})
\IF{$a_t^\star=\textit{request}(v_j)$}
\STATE Reacquire $v_j$, reconstruct all affected features, and rerun PGIK-Fusion
\STATE \rev{Release the updated prediction directly without a second gate}
\ENDIF
\RETURN Final label or handoff status, together with the action record
\end{algorithmic}
\end{algorithm}

\rev{The evaluated policy is single-step: it permits at most one sensor request, and the post-request prediction is released without another accept/request/handoff decision.}

Comparator policies isolate which part of the decision rule provides value. Confidence-only and quality-only gates use thresholds selected by validation expected cost, whereas the action-rate-matched random policy preserves aggregate accept, request, and handoff counts while randomizing their allocation and requested sensor. The Oracle policy accesses realized restoration outcomes only to define a retrospective lower bound. Unsafe acceptance is the error fraction among accepted windows. Risk--coverage follows selective classification \cite{elyaniv_2010_selective}, with calibrated confidence used as a post-hoc ranking score \cite{cattelan_2024_confidence}.

\subsection{Datasets and Evaluation Protocols}

The primary protocol evaluates person-independent logistics activity recognition. The MotionMiners IMU subset of LARa version 03 contains three IMUs, 27 channels, and 100-Hz sampling \cite{niemmann_2023_lara_v03}. We use 200-sample (2-s) windows with 100-sample stride and majority labels. Removing `None' and windows below 0.5 purity yields 21,952 windows in seven classes; \emph{Handling centre} and \emph{Synchronisation} constitute approximately 57\% and 1.6\%, respectively.

\prevrev{\rev{The MotionMiners IMU modality used in this study is available for S07--S14 but not for S01--S06 \cite{niemmann_2023_lara_v03}.} Modality availability therefore defines the eight-subject cohort. To prevent windows from the same person appearing across development and test partitions, S07--S10 (11,424 windows) are used for model fitting, S11--S12 (3,816 windows) for early stopping, fusion, calibration, and policy fitting, and S13--S14 (6,712 windows) for final evaluation. The resulting person-held-out estimates apply to the MotionMiners subset rather than to the full LARa subject population. Subject-level window and class counts are reported in Supplementary Table~S3.}

\rev{The auxiliary protocol evaluates the framework within the recorded event under equipment-sensing semantics.} MDWTBM-LS contains synchronized load, strain, and vibration measurements around one recorded wind-turbine blade lightning event \cite{li_2025_mdwtbm}. Each model uses a 30-min history to forecast the maximum Normal, Transition, or Anomaly state in the following 60~min. Models are retrained with seeds 2026, 2027, and 2028, and all seed-specific selection quantities are fitted on the corresponding validation partition. In this configuration, validation mean true-class NLL reduction selects the candidate modality for each degradation; validation-trained risk and recovery gates then allocate accept, request, and handoff actions. Label intervals, window construction, splitting, physical descriptors, gate configurations, and seven modality-degradation conditions are provided in Supplementary Section S1.

\subsection{Baselines and Implementation Details}

\prevrev{The comparison set includes feature-based trees \cite{breiman_2001_random_forests,chen_2016_xgboost,ke_2017_lightgbm}; recurrent, convolutional, and self-attention encoders \cite{hochreiter_1997_lstm,cho_2014_gru,bai_2018_tcn,vaswani_2017_attention}; Gramian angular field (GAF), recurrence plot (RP), and short-time Fourier transform (STFT) models \cite{wang_2015_imaging_timeseries}; Sensor-GNN, TimesNet, ModernTCN, and TimeMIL \cite{kipf_2017_gcn,wu_2023_timesnet,luo_2024_moderntcn,chen_2024_timemil}; and probability fusion. Transformed-domain and graph models are evaluated without XGBoost fusion. Supplementary Section S4.1 gives the configurations.}

All LARa neural models, including the temporal branch of PGIK-Fusion, use class-weighted cross-entropy, AdamW with learning rate $2\times10^{-3}$ and weight decay $10^{-4}$, batch size 128, at most 50 epochs, and early stopping after 10 epochs without improvement in validation Macro-F1. The shared optimization budget controls training differences across neural baselines, while early stopping retains the checkpoint with the highest validation Macro-F1 for fusion, intervention, and efficiency analyses. Random-number generators in Python, NumPy, and PyTorch are initialized with the designated experimental seed. The PGIK-Fusion multi-seed experiment uses seeds 2026, 2027, and 2028.

Accuracy, balanced accuracy, and Macro-F1 are computed on held-out test windows. Structural scale counts trainable neural parameters and boosted-tree nodes. Experiments were run on a Mac mini with a 10-core Apple M4 processor and 16~GB unified memory under macOS 26.2. Neural-network training used the PyTorch Metal Performance Shaders (MPS) backend where supported; no Compute Unified Device Architecture (CUDA) device was used. All reported latency measurements placed models and tensors on the central processing unit (CPU) and did not invoke MPS. Single-window latency is measured with one CPU thread and batch size one after at least 20 warm-up calls, using 300 timed calls. PGIK-Fusion timing includes vectorized physical feature construction, XGBoost prediction, TorchScript 1D-CNN inference, and probability fusion. Its reported value is the median of seven trials, each preceded by 30 warm-up calls.

\section{Results and Discussion}

\subsection{Perception Comparison and Backbone Selection}

\prevrev{Table~\ref{tab:baseline} compares all models under the S07--S10/S11--S12/S13--S14 protocol. Selection uses validation Macro-F1, and every reported value is computed on held-out S13--S14.}

\begin{table}[!t]
\caption{Test performance, structural scale, and single-window CPU latency under the fixed subject protocol. Acc. and Bal. Acc. denote accuracy and balanced accuracy; scale counts trainable parameters, boosted-tree nodes, or both for fusion models.\label{tab:baseline}}
\centering\scriptsize
\begin{tabular*}{\columnwidth}{@{\extracolsep{\fill}}lrrrrr@{}}\toprule
Model & Acc. & \shortstack{Bal.\\Acc.} & \shortstack{Macro-\\F1} & Scale & \shortstack{Time\\(ms)} \\\midrule
PGIK-Fusion & \textbf{0.768} & \textbf{0.664} & \textbf{0.696} & 296,591 & 0.520 \\
XGBoost+1D-CNN & 0.743 & 0.664 & 0.676 & 297,995 & 0.278 \\
XGBoost & 0.753 & 0.630 & 0.674 & 119,940 & 0.089 \\
LightGBM & 0.746 & 0.618 & 0.660 & 102,196 & 0.460 \\
STFT-CNN & 0.739 & 0.615 & 0.630 & 169,543 & 0.195 \\
TimesNet & 0.627 & 0.654 & 0.596 & 314,183 & 1.632 \\
TimeMIL & 0.588 & 0.658 & 0.585 & 240,728 & 0.256 \\
ModernTCN & 0.659 & 0.599 & 0.568 & 46,471 & 2.017 \\
Transformer & 0.647 & 0.632 & 0.593 & 152,935 & 0.672 \\
GRU & 0.587 & 0.632 & 0.574 & 92,551 & 2.697 \\
1D-CNN & 0.546 & 0.646 & 0.546 & 178,055 & 0.198 \\
Sensor-GNN & 0.603 & 0.560 & 0.537 & 177,975 & 0.306 \\
RP-CNN & 0.431 & 0.476 & 0.425 & 100,679 & 0.511 \\
tCNN-softmax & 0.621 & 0.621 & 0.567 & 71,303 & 0.174 \\
tCNN-semantic & 0.580 & 0.645 & 0.566 & 72,538 & 0.179 \\
ResNeXt-1D & 0.583 & 0.633 & 0.564 & 646,027 & 0.875 \\\bottomrule\end{tabular*}
\end{table}

PGIK-Fusion achieved the highest accuracy (0.768) and Macro-F1 (0.696), exceeding XGBoost--1D-CNN by 2.49 and 2.00 percentage points, respectively. Both models had a rounded balanced accuracy of 0.664, which motivates the class-level analysis below. The strongest modern standalone model, TimesNet, reached 0.596 Macro-F1. Overall, fusion models and boosted trees formed the leading group, and performance did not increase monotonically with architectural recency. Historical LARa results in Section~II-A use different modalities, partitions, and metrics and are not numerically comparable with the present subject-held-out IMU protocol.

\begin{figure}[!t]
\centering
\includegraphics[width=0.92\columnwidth]{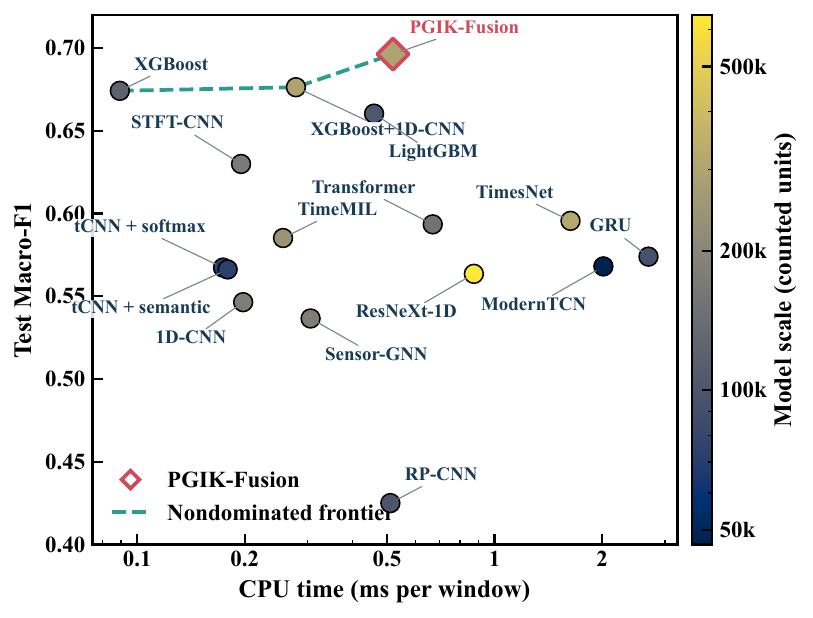}
\caption{Test Macro-F1 versus single-window CPU latency on S13--S14; marker color denotes structural scale, and the dashed line joins nondominated points.\label{fig:efficiency}}
\end{figure}

\prevrev{Deployment suitability depends jointly on predictive quality and execution cost. Fig.~\ref{fig:efficiency} therefore relates test Macro-F1 to single-window CPU latency for all 16 models; color denotes structural scale and the dashed line joins operating points for which neither metric can be improved without worsening the other.}

PGIK-Fusion occupied the highest-performance endpoint of the nondominated frontier. Relative to XGBoost--1D-CNN, its 2.00-percentage-point Macro-F1 gain increased median latency from 0.278 to 0.520~ms per window; both remained below 1~ms. XGBoost was faster at 0.089~ms but 2.20 percentage points lower in Macro-F1. The comparison therefore supports PGIK-Fusion when the additional 0.242~ms is acceptable, while identifying XGBoost as the lower-latency operating point. Structural scale and latency were not monotonically related, as illustrated by the compact but slower ModernTCN.

\prevrev{Table~\ref{tab:fusion-backbone} compares compact temporal backbones using validation-selected XGBoost weights of 0.50, 0.45, 0.90, and 0.65 for GRU, ModernTCN, Transformer, and 1D-CNN. Their combined structural scales are 212,491, 166,411, 272,875, and 297,995 counted units, respectively. TimesNet remains standalone because its fusion would contain 434,123 counted units. Latency is the seven-trial median for joint branch inference under the batch-one, single-thread CPU protocol in Section~III-E.}

\begin{table}[!t]
\caption{Test performance and single-window CPU latency of compact fusion backbones on S13--S14.\label{tab:fusion-backbone}}
\centering\footnotesize
\setlength{\tabcolsep}{2.0pt}
\begin{tabular}{@{}lrrrr@{}}\toprule
Fusion backbone & Accuracy & Bal. Acc. & Macro-F1 & \shortstack{Time\\(ms)} \\\midrule
XGBoost+GRU & 0.701 & 0.654 & 0.638 & 2.923 \\
XGBoost+ModernTCN & 0.744 & 0.647 & 0.660 & 2.232 \\
XGBoost+Transformer & \textbf{0.751} & 0.630 & 0.672 & 0.777 \\
XGBoost+1D-CNN & 0.743 & \textbf{0.664} & \textbf{0.676} & \textbf{0.278} \\\bottomrule\end{tabular}
\end{table}

XGBoost--1D-CNN provided the highest Macro-F1 (0.676) and balanced accuracy (0.664). XGBoost--Transformer was 0.004 lower in Macro-F1 and required 2.8 times the latency, while XGBoost--ModernTCN reduced scale by 44\% but lost 1.6 percentage points of Macro-F1 and required 8.0 times the latency. The 1D-CNN therefore provided the strongest performance--latency balance among the tested fusion backbones.

\subsection{Physical Evidence, Class Effects, and Efficiency}

\prevrev{Table~\ref{tab:physical-ablation} adds vector magnitude, kinematic difference, and directional consistency separately before evaluating their joint combination. This design distinguishes descriptor-specific physical information from a generic increase in feature dimension.}

\begin{table}[!t]
\caption{Test performance of the physical-feature ablation on S13--S14.\label{tab:physical-ablation}}
\centering\footnotesize
\setlength{\tabcolsep}{3.1pt}
\begin{tabular}{@{}lrrr@{}}\toprule
Feature configuration & Accuracy & Bal. Acc. & Macro-F1 \\\midrule
Raw fusion & 0.743 & 0.664 & 0.676 \\
+ Vector magnitude & 0.755 & 0.678 & 0.690 \\
+ Kinematic difference & 0.743 & \textbf{0.698} & 0.685 \\
+ Directional consistency & 0.755 & 0.684 & 0.694 \\
Full PGIK-Fusion & \textbf{0.768} & 0.664 & \textbf{0.696} \\\bottomrule
\end{tabular}
\end{table}

\prevrev{Vector magnitude, kinematic difference, and directional consistency increased Macro-F1 by 1.34, 0.82, and 1.73 percentage points over raw fusion, respectively. Directional consistency recovered approximately 86\% of the full 2.00-percentage-point gain, while the joint combination added only 0.27 percentage points beyond that variant. Conversely, the kinematic-only variant attained the highest balanced accuracy (0.698), 3.43 percentage points above the full model. The descriptor effects were therefore neither additive nor uniform across metrics, motivating the classwise analysis below.}

\begin{figure}[!t]
\centering
\includegraphics[width=0.92\columnwidth]{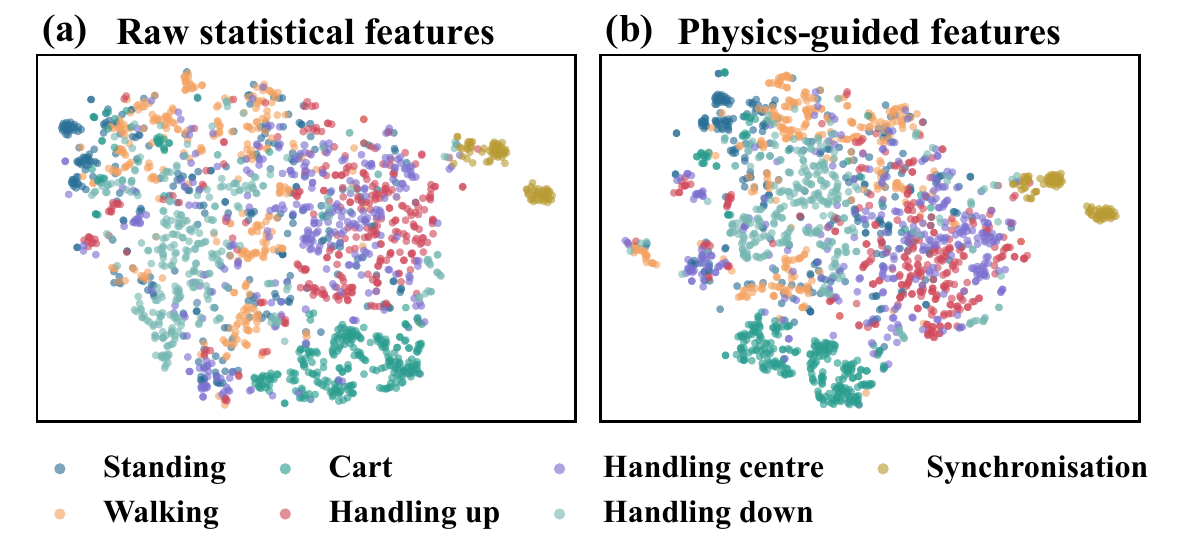}
\caption{Deterministic t-SNE projections of sampled test windows using (a) raw statistical features and (b) physics-guided structured features.\label{fig:tsne}}
\end{figure}

To examine how physical augmentation reorganized the structured feature space, Fig.~\ref{fig:tsne} compares deterministic t-distributed stochastic neighbor embedding (t-SNE) projections of the same sampled test windows before and after adding the invariant--kinematic descriptors. Up to 300 windows per class are sampled with seed 2026; each representation is standardized, reduced to 30 principal components, and embedded with perplexity 30 over 1200 iterations.

The physics-guided projection formed locally tighter \emph{Cart} and \emph{Synchronisation} regions, whereas the handling activities remained substantially overlapped. This geometry is consistent with selective class separation rather than uniform disentanglement. Because neighborhood projections do not quantify classification error, the interpretation is treated as descriptive and is evaluated against the confusion and recall evidence below.

\begin{figure}[!t]
\centering
\includegraphics[width=0.92\columnwidth]{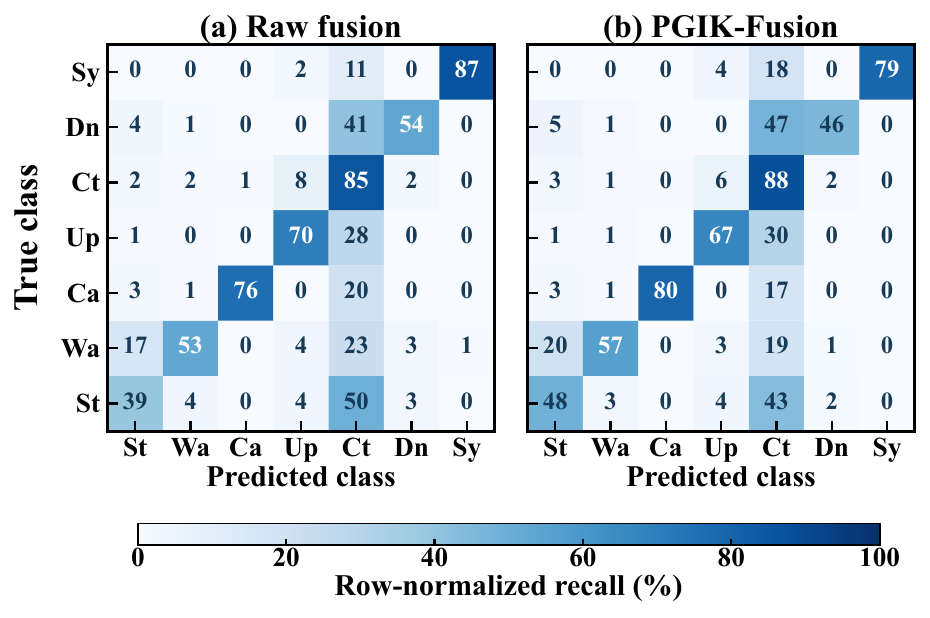}
\caption{Row-normalized test confusion matrices for (a) raw XGBoost--1D-CNN fusion and (b) full PGIK-Fusion. St, Wa, Ca, Up, Ct, Dn, and Sy denote Standing, Walking, Cart, Handling up, Handling centre, Handling down, and Synchronisation, respectively.\label{fig:confusion}}
\end{figure}

To determine whether the projected changes translated into classwise decisions, Fig.~\ref{fig:confusion} reports the complete row-normalized confusion matrices for raw fusion and PGIK-Fusion.

\begin{figure}[!t]\centering\includegraphics[width=0.92\columnwidth]{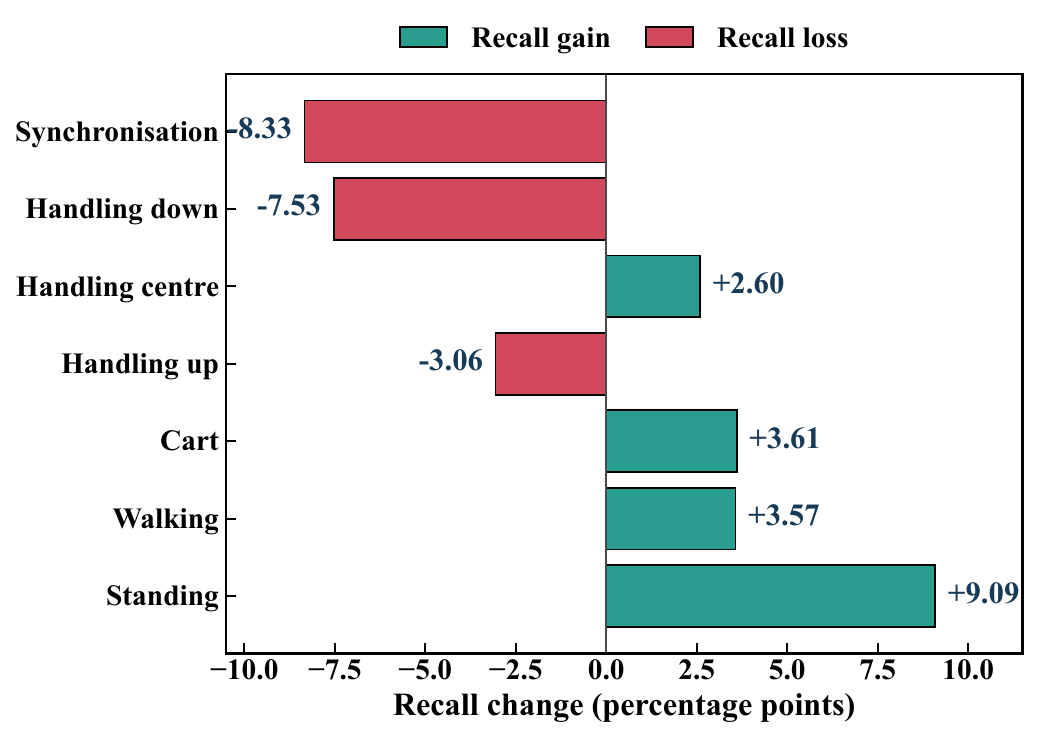}
\caption{Per-class test recall change from raw fusion to full PGIK-Fusion.\label{fig:recall-delta}}\end{figure}

\prevrev{The matrices show that physical augmentation redistributed rather than uniformly reduced errors. Fig.~\ref{fig:recall-delta} summarizes the diagonal differences as signed recall changes.}

PGIK-Fusion increased recall for \emph{Standing}, \emph{Walking}, \emph{Cart}, and \emph{Handling centre} by 9.09, 3.57, 3.61, and 2.60 percentage points, respectively. Recall decreased by 3.06 percentage points for \emph{Handling up}, 7.53 percentage points for \emph{Handling down}, and 8.33 percentage points for \emph{Synchronisation}. The 17.42-percentage-point span between the largest gain and loss shows substantial class-level redistribution beneath the 2.00-percentage-point aggregate gain. \prevrev{\rev{This redistribution motivates classwise reliability analysis alongside aggregate Macro-F1. The reported safety improvement is aggregate: the gate routed 83.9\% of \emph{Standing} and 84.85\% of \emph{Handling down} class-condition windows to request or handoff. Direct acceptance remained unreliable for these classes, with erroneous predictions in all accepted \emph{Standing} instances and 77.84\% of accepted \emph{Handling down} instances. Relative to releasing every factual prediction, erroneous accepted outputs across all class-condition windows fell from 76.3\% to 16.1\% for \emph{Standing} and from 65.23\% to 11.79\% for \emph{Handling down}. For \emph{Synchronisation}, routing 57.82\% of instances to request or handoff reduced this exposure from 43.54\% to 4.25\%. Protection for \emph{Standing} and \emph{Handling down} therefore arose mainly from routing rather than reliable direct acceptance, and the classwise reliability imbalance persisted; Supplementary Table~S6 reports the complete action and error profile.}}

Across seeds 2026, 2027, and 2028, PGIK-Fusion yielded a test Macro-F1 of $0.694\pm0.006$, and all three runs remained above the raw XGBoost--1D-CNN fusion. Per-seed metrics are provided in Supplementary Fig. S1. Table~\ref{tab:baseline} shows that PGIK-Fusion contains 118,536 boosted-tree nodes and 178,055 trainable 1D-CNN parameters. Under the stated CPU protocol, vectorized feature construction, XGBoost prediction, TorchScript 1D-CNN inference, and probability fusion together required a median of 0.520 ms per window. \prevrev{The post-perception gate comprises 70 fitted coefficients across the logistic risk model and ridge utility model. Under the same batch-one, single-thread protocol, including construction and evaluation of all three candidate utility inputs and the final cost comparison, its median latency was 0.179 ms per window, or 34.4\% of the 0.520-ms perception latency; the measured perception-plus-policy path was therefore 0.698 ms.} Fig.~\ref{fig:efficiency} places the perception measurement in the performance--latency distribution. Structural scale and latency characterize complementary aspects of deployment: the former combines heterogeneous neural-parameter and tree-node counts and is not a proxy for memory consumption or floating-point operations. \prevrev{Supplementary Table~S5 reports the timing distribution.}

\prevrev{The directional-consistency variant recovered most of the aggregate gain, but class effects remained nonuniform. The submillisecond perception-plus-policy path supports a separate safety layer for unreliable classes and sensor states.}

\subsection{Calibrated Evidence and Selective Risk}

\begin{figure}[!t]\centering\includegraphics[width=0.92\columnwidth]{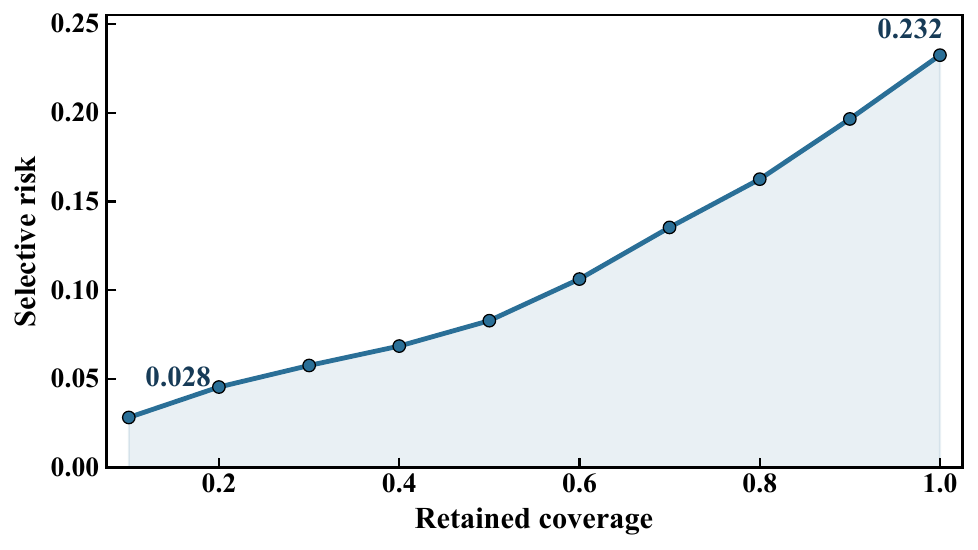}
\caption{Selective test risk as retained coverage increases for temperature-scaled PGIK-Fusion probabilities.\label{fig:risk}}\end{figure}

All calibration and selection parameters were fitted on S11--S12. Fig.~\ref{fig:risk} ranks test windows by calibrated confidence and reports selective risk as coverage increases.

\prevrev{Temperature scaling selected $T=1.45$. Selective risk increased from 0.028 at 10\% coverage to 0.083 at 50\% and 0.232 at full coverage, corresponding to accepted accuracies of 97.2\%, 91.7\%, and 76.8\%. Increasing coverage from 10\% to 50\% therefore tripled risk while adding 40 percentage points of automation; full coverage added a further 14.9 percentage points of risk. At $\alpha=0.10$, split conformal prediction achieved 0.929 empirical coverage, 2.88 percentage points above the nominal target, with a mean set size of 1.80. The risk--coverage curve and conformal sets characterize confidence-ranked and set-valued uncertainty, respectively, for the evaluated split.}

\subsection{Counterfactual Sensor Recovery and Safety Actions}

\begin{figure}[!t]\centering\includegraphics[width=\columnwidth]{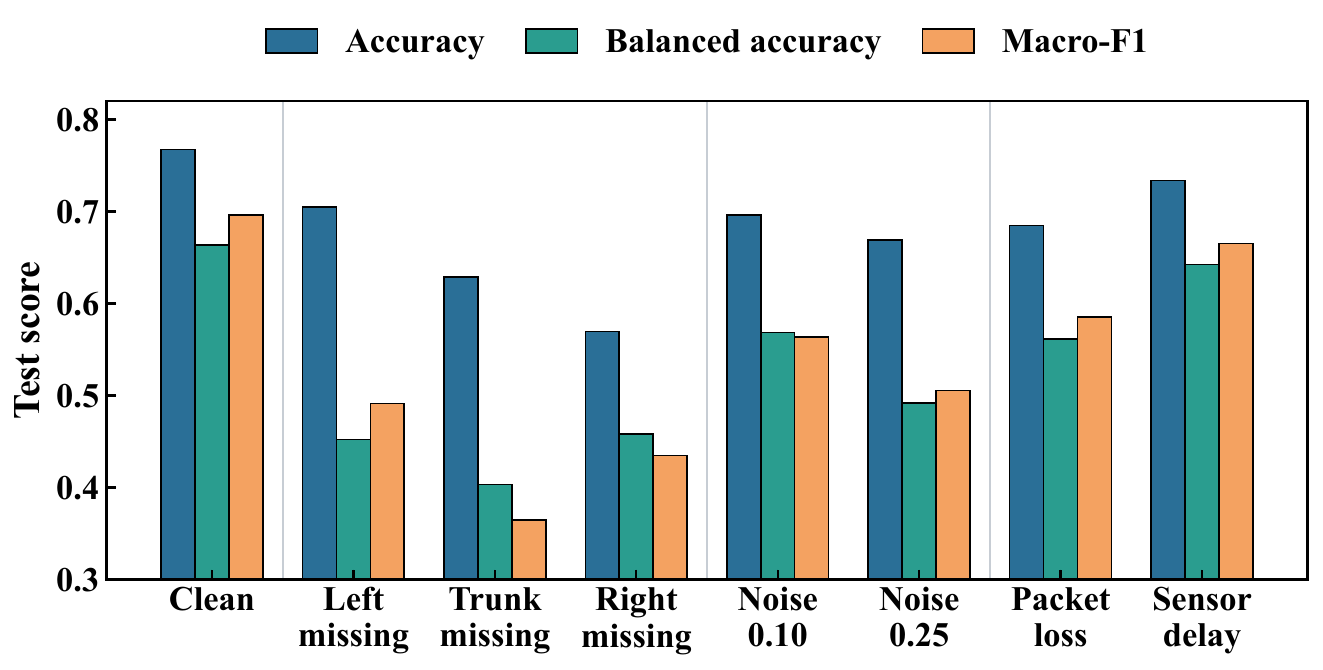}
\caption{Test recognition performance under controlled missingness, noise, packet-loss, and delay interventions.\label{fig:robustness}}\end{figure}

All structured and 1D-CNN inputs were recomputed after each controlled intervention. Fig.~\ref{fig:robustness} compares complete IMU loss, distributed noise, packet loss, and temporal delay.

\prevrev{Missing left, trunk, and right IMUs reduced Macro-F1 from 0.696 to 0.492, 0.365, and 0.435, corresponding to losses of 20.48, 33.17, and 26.16 percentage points. Trunk loss removed 47.6\% of the clean Macro-F1 and was the most damaging intervention. Moderate noise, strong noise, packet loss, and right-IMU delay retained 0.564, 0.506, 0.586, and 0.665, respectively. The 3.09-percentage-point delay loss was less than one sixth of the smallest complete-sensor loss, confirming that recognition severity differed markedly by fault type.}

\begin{figure}[!t]\centering\includegraphics[width=\columnwidth]{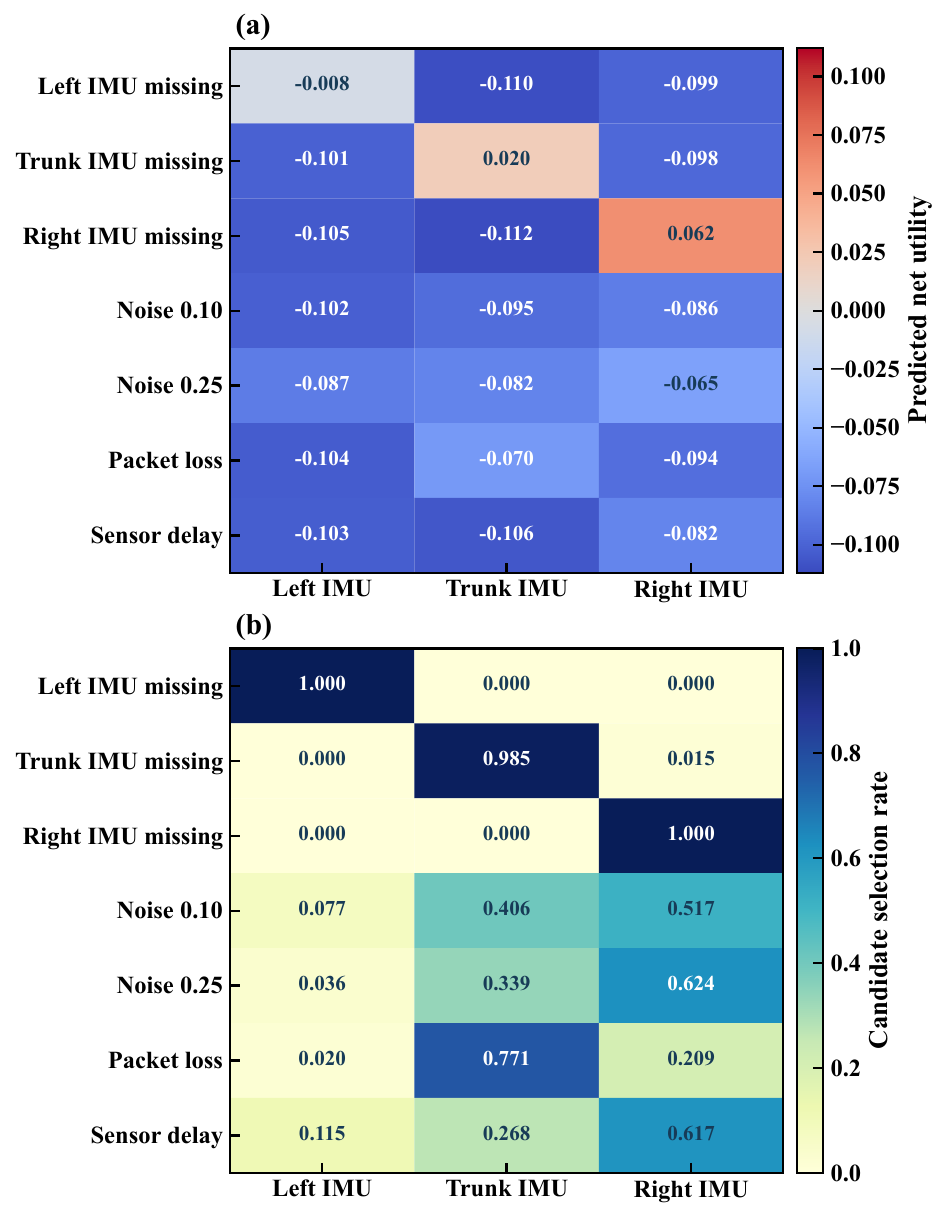}
\caption{Window-specific candidate sensor evaluation under seven LARa degradations: (a) mean predicted benefit minus request cost; (b) fraction assigned to each candidate before the final action comparison.\label{fig:utility}}\end{figure}

Recognition loss alone does not establish whether an additional sensor should be requested. Fig.~\ref{fig:utility} therefore evaluates candidate-specific net utility before the final action comparison.

Restoring the affected sensor produced positive mean net utility for trunk and right-IMU loss (0.021 and 0.062), whereas restoring unaffected sensors produced negative values. The selector chose the matching sensor for 98.5\% of trunk-loss windows and all left- and right-loss windows. Under distributed noise, packet loss, and delay, all mean utilities were negative after request cost. The highest-ranked candidate was therefore not necessarily preferable to acceptance or handoff. Degradation severity and recoverability were distinct: the most corrupted signal was not always cost-effective to reacquire.

\prevrev{Held-out restoration outcomes evaluate the utility ranking against an imbalanced signed target. Among 140,952 window--candidate pairs, 6,348 (4.50\%) were beneficial, 132,453 (93.97\%) were neutral, and 2,151 (1.53\%) were harmful. \rev{The three candidates for each of 46,984 degraded window--condition instances account for these pairs. Of the 46,984 instances, 5,888 (12.53\%) contained at least one corrective restoration; the selected candidate corrected 5,018, yielding an opportunity-capture rate of 85.22\%.} Predicted benefit had a mean absolute error of 0.087 and a Pearson correlation of 0.219 with realized 0--1 loss reduction. The predominance of neutral targets contributes to the low mean absolute error, while the modest correlation indicates limited global linear agreement. The opportunity-capture rate quantifies candidate ranking within the corrective-opportunity subset. The estimand remains the controlled frozen-model response defined in Section~III-C. Mean regret relative to the retrospectively best candidate was 0.048 loss units per window. Supplementary Table~S4 reports the target distribution. Together, Figs.~\ref{fig:robustness} and~\ref{fig:utility} separate fault severity from recoverability and connect both quantities to the finite action space.}

\subsection{Counterfactual Policy Comparison and Agent Decisions}

\begin{figure}[!t]\centering\includegraphics[width=0.94\columnwidth]{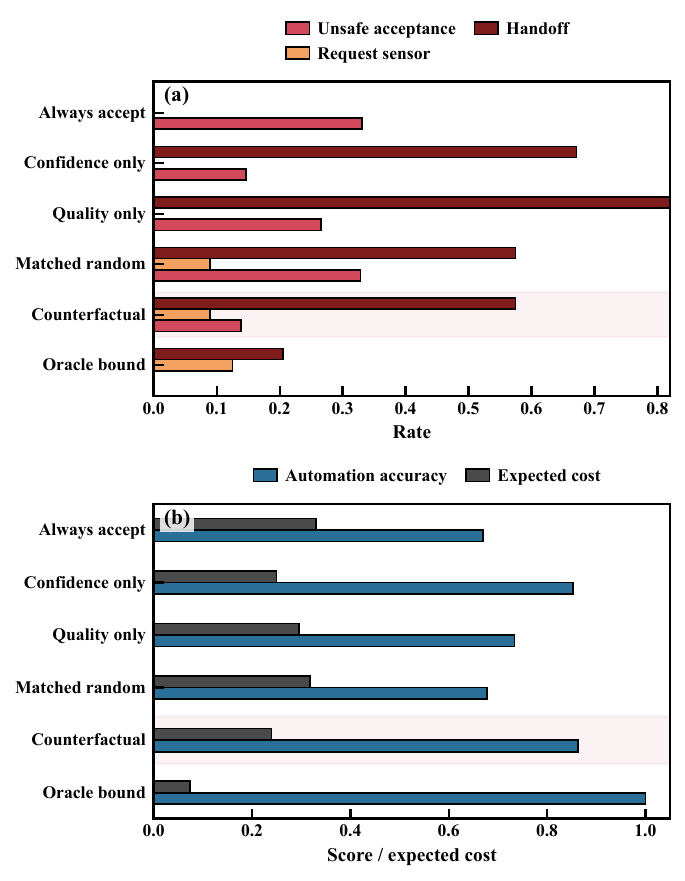}
\caption{Safety-policy comparison across seven degraded test conditions: (a) unsafe acceptance and action rates; (b) automated accuracy and expected cost.\label{fig:policy}}\end{figure}

\prevrev{To test whether candidate-specific utility improves the allocation of a fixed action budget, Fig.~\ref{fig:policy} compares the counterfactual policy with always accepting, confidence-only and quality-only gating, action-rate-matched random allocation, and the Oracle lower bound. The random policy preserves the counterfactual policy's aggregate accept, request, and handoff rates while randomizing their allocation and requested sensor. Differences from this comparator therefore reflect where the policy acts rather than how often it acts. Supplementary Table~S7 provides the numerical values.}

Across seven degraded conditions, always accepting produced 33.03\% unsafe acceptance. The counterfactual policy reduced this rate to 13.84\%, an absolute decrease of 19.19 percentage points and a relative decrease of 58.1\%. It accepted 33.64\%, requested one sensor for 8.89\%, and handed off 57.47\% of windows. At the same action rates, the action-rate-matched random policy produced 32.86\% unsafe acceptance, 0.678 automation accuracy, and 0.319 expected cost. Candidate-specific allocation reduced unsafe acceptance by 19.02 percentage points, increased automation accuracy by 18.51 percentage points, and lowered cost by 24.7\% to 0.240. The gain therefore arose from action allocation rather than a higher handoff rate.

Validation cost minimization selected no request actions for either scalar comparator. The confidence-only policy obtained 14.68\% unsafe acceptance, 0.853 automation accuracy, and 0.250 expected cost. Quality-only gating accepted 14.3\% of windows and obtained 26.6\% unsafe acceptance, 0.734 automation accuracy, and 0.295 expected cost, showing that degradation metadata alone did not resolve class uncertainty. Relative to confidence-only gating, the counterfactual policy reduced unsafe acceptance by 0.83 percentage points, increased automation accuracy by 0.93 percentage points, and reduced cost by 4.0\%, while retaining targeted reacquisition. Across request costs of 0.05--0.25, the request rate fell from 22.37\% to zero at costs of at least 0.20; expected cost ranged from 0.236 to 0.242.

The counterfactual policy remained 0.166 above the Oracle cost of 0.074. Observed sources of this gap include candidate-selection regret and errors in the estimated risk and benefit. Its performance therefore remained between action-rate-matched random allocation and the retrospective Oracle lower bound.

\begin{figure}[!t]\centering\includegraphics[width=0.94\columnwidth]{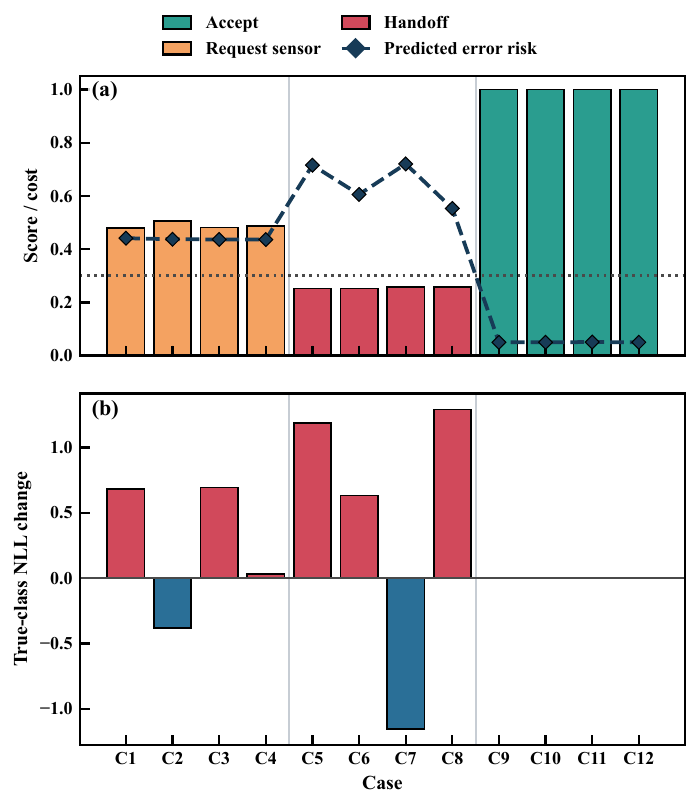}
\caption{Illustrative one-step decisions under IoT sensor degradation: (a) confidence, predicted error risk, and selected action; (b) retrospective change in true-class NLL.\label{fig:cases}}\end{figure}

Aggregate policy metrics do not show how an individual sensing decision is formed. Fig.~\ref{fig:cases} therefore traces representative decisions from observed risk to action. Retrospective true-class NLL is used only to evaluate the selected action.

Cases C1--C4 are right-IMU-loss windows for which the predicted risk reduction exceeded request cost. Cases C5--C8 were handed off because their predicted residual risk remained high. Cases C9--C12 combined low predicted error risk with no cost-effective request and were accepted. The cases illustrate the finite, single-step decisions defined in Section~III-A without supplying post-request outcomes to the policy. The policy comparison and cases jointly support the finite sensing-agent formulation by linking pre-request evidence to actions over a defined set of independently restorable sensors.

\subsection{\rev{Auxiliary Study in Wind-Turbine Blade Monitoring}}

\rev{To examine cross-domain re-instantiation within the recorded event,} the models were retrained under the MDWTBM-LS block-partition protocol with seeds 2026, 2027, and 2028. The auxiliary task forecasts the maximum Normal, Transition, or Anomaly state in the next 60 min from wind-turbine blade load, strain, and vibration histories. Table~\ref{tab:mdwtbm} compares the retrained perception models.

\begin{table}[!t]
\caption{Three-seed test performance for 60-min-ahead wind-turbine blade state forecasting on MDWTBM-LS. Entries are mean $\pm$ standard deviation over seeds 2026, 2027, and 2028; \prevrev{all partitions are derived from the same recorded lightning event.}\label{tab:mdwtbm}}
\centering\scriptsize
\setlength{\tabcolsep}{1.7pt}
\begin{tabular}{@{}lccc@{}}\toprule
Model & Acc. & Bal. Acc. & Macro-F1 \\\midrule
XGBoost & $0.935\pm0.050$ & $0.904\pm0.063$ & $0.924\pm0.052$ \\
XGBoost+1D-CNN & $0.971\pm0.023$ & $0.946\pm0.026$ & $0.963\pm0.020$ \\
PGIK-Fusion & $\mathbf{0.978\pm0.022}$ & $\mathbf{0.964\pm0.031}$ & $\mathbf{0.975\pm0.022}$ \\\bottomrule
\end{tabular}
\end{table}

PGIK-Fusion reached $0.975\pm0.022$ Macro-F1, compared with $0.963\pm0.020$ for XGBoost--1D-CNN and $0.924\pm0.052$ for XGBoost. The paired changes over raw fusion were +1.77, +1.73, and +0.00 percentage points, yielding a mean increase of 1.17 percentage points. Two seeds improved and one tied; the gain was therefore positive on average but not uniform across seeds.

Calibration effects differed by metric: temperature scaling reduced mean ECE and Brier score but increased test NLL. At $\alpha=0.10$, conformal coverage was $0.909\pm0.110$. Supplementary Fig.~S3(a) reports the complete calibration and risk--coverage evidence. Removing load, strain, and vibration reduced Macro-F1 by 32.25, 38.50, and 37.70 percentage points, respectively. High clean performance therefore coexisted with substantial modality dependence; complete condition-level results are provided in Supplementary Fig.~S2.

\begin{figure}[!t]\centering
\includegraphics[width=0.94\columnwidth]{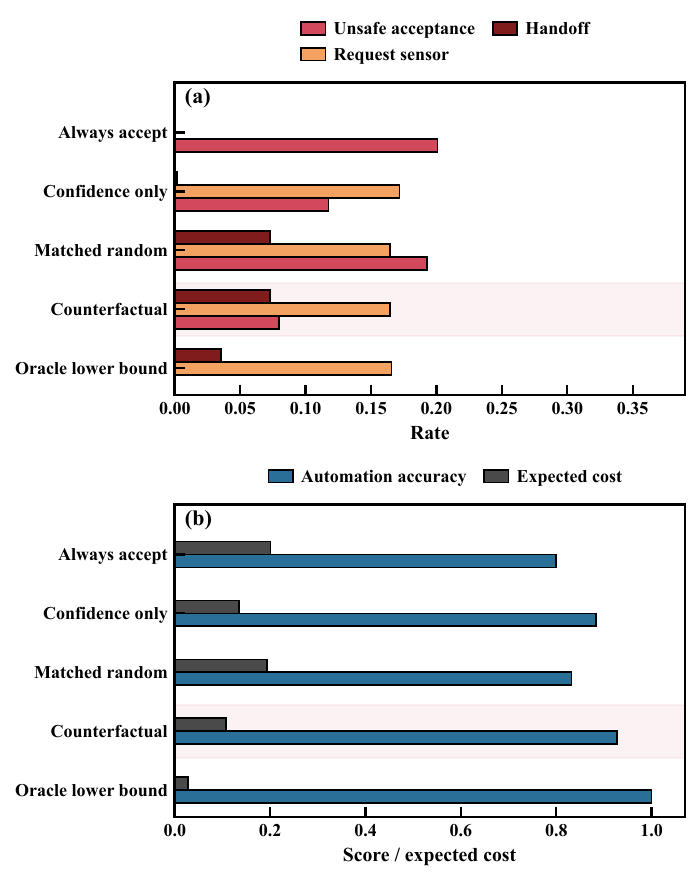}
\caption{Finite-action policy comparison across seven degraded MDWTBM-LS sensing conditions: (a) unsafe acceptance, sensor request, and handoff rates; (b) automation accuracy and expected cost.\label{fig:mdwtbm-policy}}
\end{figure}

To compare decision policies under these degraded conditions, Fig.~\ref{fig:mdwtbm-policy} presents the mean unsafe acceptance, request and handoff rates, automation accuracy, and expected cost across the three runs. Acceptance is the complement of request and handoff and is therefore omitted as a redundant series. The action-rate-matched random policy uses the same aggregate action rates as the counterfactual policy, and the Oracle is retained as a retrospective lower bound. Supplementary Table~S8 lists the corresponding means and standard deviations, including the complementary acceptance rate.

The counterfactual policy reduced unsafe acceptance from $0.201\pm0.030$ under always accepting to $0.080\pm0.057$, a 60.2\% relative reduction. Against the action-rate-matched random policy, it reduced unsafe acceptance by 11.28 percentage points, increased automation accuracy by 9.57 percentage points, and lowered expected cost by 44.5\%, from $0.194\pm0.024$ to $0.108\pm0.033$. Relative to confidence-only gating, unsafe acceptance fell by 3.76 percentage points and cost by 19.7\%, while request rates were similar (0.165 versus 0.172). The counterfactual policy also handed off 7.3\% of windows.

\prevrev{Condition-specific results in Supplementary Fig.~S2 and uncertainty and case analyses in Supplementary Fig.~S3 complement the aggregate outcomes in Table~\ref{tab:mdwtbm} and Fig.~\ref{fig:mdwtbm-policy}. Because all training, validation, and test blocks originate from one recorded lightning event, variation across seeds measures partition and model-fitting sensitivity within that event. \rev{Four-window grouping reduces direct adjacency across partitions, but the overlapping histories and forecast horizons leave residual temporal dependence between nearby blocks.} The Macro-F1 standard deviations of 0.020--0.052 in Table~\ref{tab:mdwtbm}, together with the wider condition-level intervals in Supplementary Fig.~S2, quantify this within-event sensitivity. The 1.17-percentage-point mean perception gain and policy improvements show that the finite control abstraction can be configured with distinct sensor semantics within the recorded event. Cross-event generalization requires evaluation on independent lightning events.}

\section{Conclusion}

\prevrev{This study defines a task-bounded counterfactual evidence manager for autonomous IoT sensing. PGIK-Fusion combines physical descriptors with temporal evidence, while pre-request risk and utility connect factual and candidate-restoration states to acceptance, one sensor request, or handoff. The formulation distinguishes recognition quality, recoverability, and evidence sufficiency.}

The LARa experiments provide the primary evidence. PGIK-Fusion reached 0.696 Macro-F1, 2.00 percentage points above raw XGBoost--1D-CNN fusion. Pre-request utility captured 85.22\% of windows with a corrective restoration. At matched action rates, the counterfactual policy improved automation accuracy by 18.51 percentage points and reduced expected cost by 24.7\% relative to the action-rate-matched random policy. \rev{In the auxiliary within-event study,} PGIK-Fusion attained 0.975 Macro-F1 on MDWTBM-LS, and the counterfactual policy reduced unsafe acceptance from 20.1\% under always accepting to 8.0\%. \prevrev{These auxiliary estimates quantify partition and model-fitting sensitivity within one recorded event; cross-event generalization requires independent events.}

\prevrev{The testbeds provide complementary wearable and structural sensing evidence. Deployment requires sensor-quality metadata, reacquisition or retransmission, and action costs derived from energy, latency, bandwidth, battery state, and supervisory response.} \prevrev{The empirical evidence covers subject-held-out logistics data, one recorded wind-turbine lightning event, predefined simulated fault families and severities, and single-step actions. Unseen fault families and cross-event behavior require separate validation.} \prevrev{Long-horizon planning, continual adaptation, and autonomous physical control define extensions beyond the present single-step formulation. Within the evaluated setting, physically structured recognition, targeted evidence acquisition, and conservative deferral form an experimentally grounded perception--action loop across distinct IoT sensing configurations.}

\section*{Data Availability}
The implementation, experiment configurations, pretrained weights, and reference results are available at \url{https://github.com/shannongsr/IoTJ_counterfactual_sensor_selection}.

\bibliographystyle{IEEEtran}
\bibliography{references}

\end{document}